\documentclass[12pt]{article}
\usepackage {graphics,graphicx}
\date{}
\begin{document}
\title{Decay of photon with high as well as low energy}
\author{{\bf Indranath Bhattacharyya}\vspace{0.2cm}\\
Department of Mathematics\\
Barasat Government College\\10 K.N.C Road, Barasat (North 24 Parganas)\\ Kolkata-700124, West Bengal, India
\\E-mail :
$i_{-}bhattacharyya@hotmail.com$\\}\vskip .1in
\maketitle
\begin{abstract}
The decay of photon by the influence of magnetic field is considered. It is shown here that if the photon energy is grater than 1 MeV then photon can decay electron positron pair, but if it remains below 1 MeV then photon decays into neutrino antineutrino pair. The decay rates for both of the processes are calculated. All possible Feynman diagrams are taken into account to construct the matrix element for either of the processes. In the second process all three type of neutrinos are considered. The significance of these processes are discussed briefly.
\end{abstract}
\pagebreak
\section{Introduction}
The photon splitting phenomenon brought the attention to the
scientists and researchers. Apparently it may not be possible for
the massless photon to be decayed into the massive particles. But in
presence of strong magnetic field the photon breaks up into electron
positron pair when its energy is very high, whereas in the
comparatively low energy it gets decayed weakly into neutrino
antineutrino pair having very little mass. First time such
possibility was considered by Adler \cite{Adler}. After that many
researchers calculated this process. In 1996 the process was
considered and recalculated by Adler and Schubert \cite{Adler1}
below the pair production threshold using a variant world line path
integral approach to the Bern-Kosower formalism. Wilke and Wunner
\cite{Wilke} calculated numerically the photon splitting process and
their results confirmed the asymptotic approximations in the low as
well as high magnetic field. Some works were started to consider the
photon decay into neutrino pair. DeRaad et al. \cite{DeRaad}
considered the decay of photon into neutrino antineutrino pair to
study the significance of this process during stellar evolution. In
1998 Kuznetsov et al. \cite{Kuznetsov} calculated the probability of
the decay of photon into such neutrino pair in presence of strong
magnetic field and evaluated the contribution of the process to the
neutrino luminosity in some astrophysical circumstances. Here the calculations are carried out for the high energy as well as relatively low energy photon. Such high energetic photon may be
created in the laboratory and also during some astrophysical
phenomena like supernova. Here the low energy photon decay would
have a structural similarity with the decay of anisotropic plasma
into neutrino antineutrino pair. One must follow the usual rules of
quantum electrodynamics when the final particles are electron
positron pair, whereas in case of photon decay into neutrino
antineutrino pair the calculations are carried out in the framework
of electro-weak theory.
\section{Decay of photon into electron positron pair}
It is already stated that the photon with very high energy can be
decayed into electron positron pair. It is very interesting that
mass less photon breaks up into a pair of massive particles,
although here the energy of the photon would be so high that the
rest energy of the electron can be neglected. In presence of strong
magnetic field the photon acquires a rest mass that is considered as
the magnetic mass. The decay rate is to be calculated and in this
context the Feynman diagram, given by Figure-1, is taken into account. Let us
construct the matrix element as
$$M_{fi}^{e}=ie^{3}\varepsilon_{\mu}\Pi^{\mu\rho}(k)\frac{\overline{u}(q_{1})\gamma_{\rho}
v(q_{2})}{(q_{1}+q_{2})^{2}}\eqno{(2.1)}$$ The term
$\Pi^{\mu\rho}(k)$ present in the matrix element represents the
response tensor. It is quite clear that the term $\Pi^{\mu\rho}(k)$
arises due to the electron-positron loop present in the diagrams. It
will not have any diverging term since the gauge invariancy imposes
the restriction
$$k_{\mu}\Pi^{\mu\rho}(k)=0\eqno{(2.2)}$$

The response tensor depends on the magnetic field as well as four
momentum of the photon in plasma. If these two
parts are separated the response tensor takes the form
$$\Pi^{\mu\rho}(k)=A(k^{\mu}k^{\rho}-g^{\mu\rho}k^{2})\eqno{(2.3)}$$
Clearly this equation satisfies the gauge invariance restriction
(2.2). Note that the dimensionless quantity $A$ depends on the
magnetic field i.e. $$A=A(H)$$ where $H$ represents the magnetic
field.\\With the same argument the term $k^{\mu}$ is free from
magnetic field. The decay rate is obtained from the following
expression.
$$\tau=\frac{4\pi\mathcal{S}}{2\omega}\int\Sigma \mid M_{fi}\mid^{2}
\frac{N_{q_{1}}d^{3}q_{1}}{2q_{1}^{0}(2\pi)^{3}}
\frac{N_{q_{2}}d^{3}q_{2}}{2q_{2}^{0}(2\pi)^{3}}
\delta^{4}(k-q_{1}-q_{2})\eqno{(2.4)}$$ where $N_{i} (i=1,2)$ is
twice the mass of the corresponding fermion and $\mathcal{S}$
represents the degeneracy factor for the outgoing particles.

Calculating the term $\Sigma \mid M_{fi}^{e}\mid^{2}$ finally the decay rate is obtained as follows:
$$\tau_{e}=\alpha^{2}\frac{k^{2}}{\omega^{5}}\mid
Ak^{2} \mid^{2}\eqno{(2.5)}$$ Now we can evaluate the term $\mid
Ak^{2}\mid$ as
$$\mid Ak^{2}\mid \approx \frac{4\sqrt{\alpha}}{3\pi^{2}}(\frac{H}{H_{c}})m_{e}^{2}
\frac{\mid \overrightarrow{k}\mid}{\omega}\eqno{(2.6)}$$  Finally
the decay rate for this process can be obtained as
$$\tau_{e}
\approx4.69\times10^{9}\times(\frac{H}{H_{c}})^{2}(\frac{m_{e}}{m_{H}})^{3}
f_{e}(\eta)\hspace{1cm}sec^{-1}\eqno{(2.7)}$$ The function
$f_{e}(\eta)$ is defined as
$$f_{e}(\eta)=\eta^{5}(1-\eta^{2})\eqno{(2.8)}$$
and
$$\eta=\frac{m_{H}c^{2}}{\hbar\omega}\eqno{(2.9)}$$
where $m_{H}$ stands for the effective mass of the photon appearing
due to the presence of high magnetic field, which is given by
$$m_{H}=r(\frac{H}{H_{c}})^{\frac{1}{2}}m_{e}\eqno{(2.10)}$$
Here $r$ is a constant which makes the effective mass of the photon
(due to the presence of magnetic field) greater than the electron
mass. We can evaluate the value of $r$ as
$2(\frac{H_{c}}{H})^{\frac{1}{2}}$ in the rest frame. In the
equation (2.8) the term $\eta$ is very small which makes the
effective decay rate lower. In Table-1 we compute the decay rate of
photon in the energy range 10-100 MeV and then also in 1-100 GeV.
\section{Decay of photon into neutrino antineutrino pair}
In the calculation of decay rate it is considered the case when the
photon energy is relatively low. The Feynman diagrams for this
process are given by the Figure-2(a) and (b) indicating neutrino
emission takes place through the exchange of Z-boson and W-boson
respectively. The matrix element for this process is constructed
as

$$M_{fi}^{\nu}=-ie\frac{G_{F}}{\sqrt{2}}\varepsilon_{\mu}[g_{V}\Pi^{\mu\rho}(k)+
g_{A}\Pi_{A}^{\mu\rho}(k)]\overline{u}_{\nu}(q_{1})\gamma_{\rho}
(1-\gamma_{5})v_{\nu}(q_{2})\eqno{(3.1)}$$
 whereas $\Pi^{\mu\rho}_{A}(k)$ stands for the same, but
associated with axial vector part.n equation (3.1) the terms $g_{A}$
and $g_{V}$ are defined by\vspace{0.5cm}\\\indent
 $g_{V}=2sin^{2}\theta_{W}+\frac{1}{2}$\hspace{2cm}
 and \hspace{2cm} $g_{A}=-\frac{1}{2}$ \hspace{2cm} for \hspace{0.25cm}$\nu_{e}$
 \\\indent
$g_{V}=2sin^{2}\theta_{W}-\frac{1}{2}$\hspace{2cm} and \hspace{2cm}
$g_{A}=\frac{1}{2}$ \hspace{2cm} for \hspace{0.25cm}
$\nu_{\mu}$ and $\nu_{\tau}$ \vspace{0.5cm}\\

The term $\Pi^{\mu\rho}(k)$ would be defined in the same way as it
is defined in the equation (2.1) and the same gauge invariance
condition defined by the equation (2.2) will be applicable here.
Similarly one can find the expression for the axial vector part. It
is almost same as the equation (2.1), only difference is the
presence dimensionless term $A_{5}$ instead of $A$. The decay rate
for this process for all three type of neutrinos is calculated as
$$\tau_{\nu}=G_{F}^{2}\alpha\frac{k^{2}}{\omega}\mid
(g_{V}A+g_{A}A_{5})k^{2} \mid^{2}\eqno{(3.2)}$$ To calculate the
term $\mid (g_{V}A+g_{A}A_{5})k^{2} \mid^{2}$ the result obtained by Kennett and Melrose \cite{Kennett} in
studying the decay of anisotropic plasma may be exploited, although it is to be remembered that the decay of ordinary photon is considered. Using that result it is found
$$\mid(g_{V}A+g_{A}A_{5})k^{2}\mid \approx
\frac{\sqrt{\alpha}}{3\pi^{2}}(\frac{H}{H_{c}})m_{e}^{2} \frac{\mid
\overrightarrow{k}\mid}{\omega}\eqno{(3.3)}$$  Finally the decay
rate for this process can be obtained as
$$\tau_{\nu}\approx
3.295\times10^{-10}(\frac{H}{H_{c}})^{2}\frac{\hbar\omega}{m_{e}c^{2}}f_{\nu}(\eta)
\hspace{0.5cm}sec^{-1}\eqno{(3.4)}$$ The function $f_{\nu}(\eta)$ is
defined as
$$f_{\nu}(\eta)=\eta^{2}(1-\eta^{2})\eqno{(3.5)}$$
The equation (3.4) represents the decay rate of the splitting of
photon into neutrino pair in the C.G.S. unit.
\section{Discussion}
In the equation (2.7) the decay rate is calculated for high
energy photon. Such a high energy photon can be created in the
laboratory to verify the correctness of our calculation since the
charged electron positron pair is formed. The decay
rate is computed at a particular magnetic field ($\sim 10^{5}$ G) that can be
generated in the laboratory. It is worth noting that a higher
magnetic field can be generated in the laboratory, which depends on
the experimental setup of the magnets. In that case the decay rate
will be increased a bit. The decay of photon into electron positron
pair is possible when the energy of the photon is much higher than 1
MeV, but the result in this article shows that the decay rate will slowly diminish
with increasing the photon energy, especially when the energy goes
to the GeV range. It indicates that the decay of photon becomes
insignificant when the photon energy is extremely high. This is an
important consequence because it excludes the possibility of the
divergence of decay rate for the photon with extremely high energy.
which is not possible to verify in the laboratory. The high energy
photon decay can occur during the late stages of the stellar
evolution. The quantization of the gamma ray can provide high
energetic photon that may decay into electron positron pair. There
may also be some circumstances, for example, the cooling of highly
magnetized neutron stars and magnetars the intensity of the magnetic
field may reach to $10^{16}$ G and in presence of such super strong
magnetic field the decay rate might not be very small, but it is
quite impossible to verify such phenomenon experimentally. In case
of low energy photon the decay process yields neutrino antineutrino
pair, although the decay rate would be very small compared to the
decay of plasma in presence as well as in absence of magnetic field.
In stellar core both of the decay processes, considered here, occur
simultaneously. The weak decay of photon into neutrino antineutrino
pair may contribute to the energy loss mechanism as the mean free
path of the neutrino is longer than the stellar radii. Hence, the
photon decay is an important process in either case i.e., when the
energy of the photon is very high as well as low.\\

\noindent{\large\bf  \quad Acknowledgement :}
\vspace{0.2cm}\\ It is my pleasure to thank Prof. Probhas Raychaudhri, retired professor of Department of Applied Mathematics, University of Calcutta, India for his valuable suggestions and
guidance to prepare this manuscript.

\pagebreak

\begin{table}
\begin{tabular}{|c|c|c|c|}\hline
E (MeV)&$\tau_{e}$ $(sec^{-1})$&E (GeV)&$\tau_{e}$
$(sec^{-1})$\\\hline
$10$ & $3.48 \times 10^{-7}$&$1$ & $2.15 \times 10^{-22}$\\
$20$ & $1.09 \times 10^{-8}$&$10$ & $2.15 \times 10^{-27}$\\
$30$ & $1.43 \times 10^{-9}$&$20$ & $6.72 \times 10^{-29}$\\
$40$ & $3.4 \times 10^{-10}$&$30$  &   $8.86 \times 10^{-30}$\\
$50$ & $1.11 \times 10^{-10}$&$40$ & $2.1 \times 10^{-30}$\\
$60$ & $4.48 \times 10^{-11}$&$50$ & $6.89 \times 10^{-31}$\\
$70$ & $2.07 \times 10^{-11}$&$60$ & $2.77 \times 10^{-31}$\\
$80$ & $1.06 \times 10^{-11}$&$70$ & $1.28 \times 10^{-31}$\\
$90$ & $5.9 \times 10^{-12}$&$80$ & $6.57 \times 10^{-32}$\\
$100$ & $3.48 \times 10^{-12}$&$90$ & $3.64 \times 10^{-32}$\\
***********&************&$100$ & $2.15 \times 10^{-32}$\\ \hline
\end{tabular}
\caption{The decay rate of the photon at the magnetic field
$H=10^{5}$ G in the energy range $10-100$ MeV and $1-100$ GeV.}
 \end{table}

\begin{figure}
\centering
\includegraphics [scale=0.4] {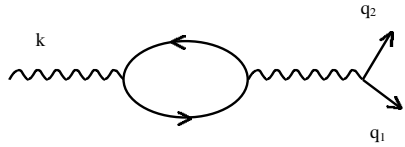}
\caption {Feynman-diagrams for photon decay in presence of magnetic
field into electron positron pair}
\end{figure}

\begin {figure}
\centering
\includegraphics [scale=0.4] {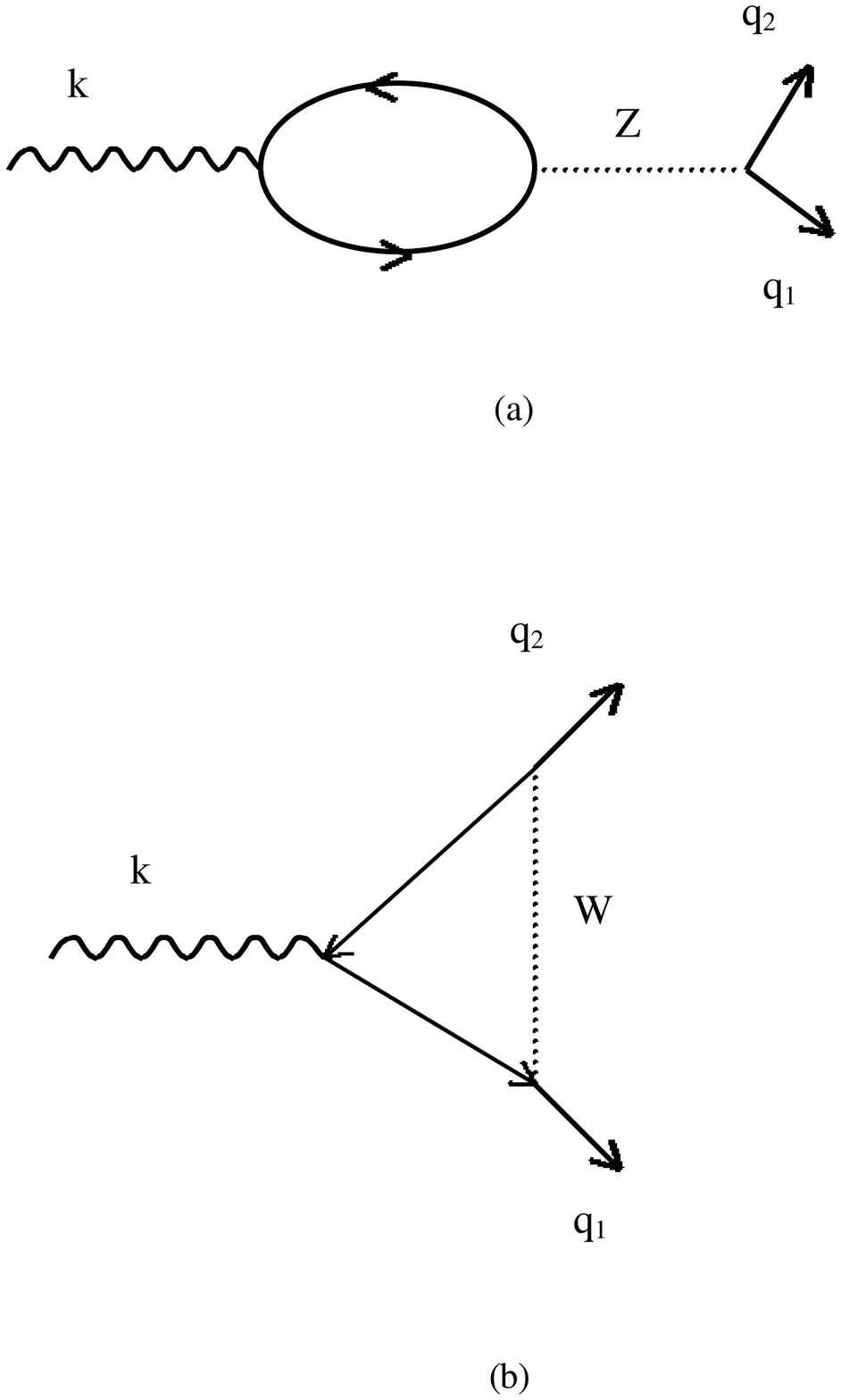}
\caption{Feynman-diagrams for photon decay in presence of magnetic
field into neutrino antineutrino pair}
\end {figure}
\end{document}